\documentclass[11pt,a4paper,twocolumn]{article}
\def \eg {e.g.}

\def \ie {i.e.}
\def \cf {cf.}

\def \rtwo {\hbox{$r_{200}$}}

\newcommand{\muG }{\mbox{$\mu$G}}

\newcommand{\rosat }{{\em ROSAT}}

\newcommand{\lofar }{LOFAR}

\newcommand{\ska }{SKA}

\newcommand{\lotss }{LoTSS}
\newcommand{\lotssE }{LOFAR Two-meter Sky Survey}

\newcommand{\sdss }{SDSS}

\usepackage{graphicx}
\usepackage{amsmath}
\usepackage{amssymb}
\usepackage{epstopdf} 
\usepackage{wrapfig}
\usepackage[bf, small]{caption}
\usepackage{geometry} 
\usepackage[]{natbib}
\setcitestyle{aysep={}}

\usepackage{hyperref}
\hypersetup{colorlinks,citecolor=blue,filecolor=black,linkcolor=red,urlcolor=black}
        
\begin{document}
\twocolumn[\center \vspace*{-3mm} 
{\bf \large Probing the magnetic field at the cluster virial radius with volume-filling radio emission}
\vspace*{4mm} \footnotesize Andrea Botteon (INAF-IRA; andrea.botteon@inaf.it) \\
\vspace*{-4mm} \footnotesize \textsl{Proceeding of the ``Cosmic Magnetism in the Pre-SKA Era'' conference (Kagoshima, Japan | May 27--31, 2024)}
\vspace*{4mm}]

\noindent
{\scriptsize {\bf Abstract.} Diffuse synchrotron emission in the form of radio halos and radio relics probe the existence of relativistic electrons and magnetic fields in galaxy clusters. These nonthermal components are generated from the dissipation of kinetic energy released by turbulence and shocks injected in the intracluster medium (ICM) during the large-scale structure formation process.
By using the deepest images ever obtained on a galaxy cluster at low-frequency (72~h LOFAR-HBA + 72~h LOFAR-LBA), in \citet{botteon22a2255} we provided an unprecedented view of the distribution of relativistic electrons and magnetic fields in the far outskirts of Abell 2255. In particular, we observed pervasive radio emission that fills the entire cluster volume and extends up to the cluster virial radius, reaching a maximum projected linear size of 5 Mpc. By combining radio and X-ray observations with advanced numerical simulations, we estimated the magnetic field and energy budget associated to turbulent motions at such large distances from the cluster center. Our results suggest an efficient transfer of kinetic energy into nonthermal components in the extremely diluted cluster outskirts.
In the past two years, the total LOFAR-HBA observation time on Abell 2255 has increased to 336 hours. The analysis of this ultra-deep dataset aims to further advance our understanding of relativistic electrons and magnetic fields in cluster peripheries.}

\subsection*{Context}

\begin{figure}[!t]
 \centering
 \includegraphics[width=\hsize,trim={3.5cm 1.5cm 16cm 2.5cm},clip]{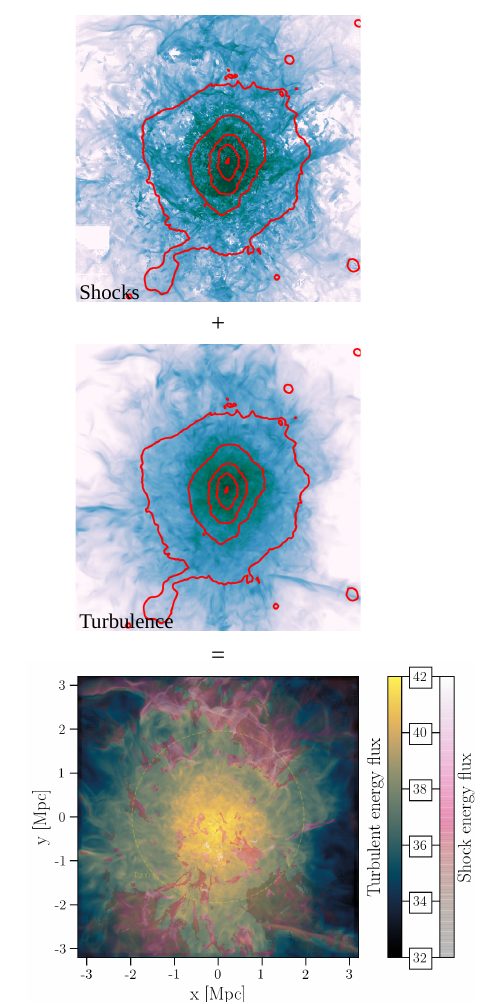}
 \caption{Simulations of the kinetic energy flux dissipated in the ICM during a cluster merger. The three panels show the contribution of shocks (\textit{top}), turbulence (\textit{center}), and shocks (with Mach number $>$2.3) and turbulence combined (\textit{bottom}). The units of the colorbar are log erg s$^{-1}$ pixel$^{-1}$. Red contours denote the X-ray emission of the simulated cluster, while the dashed yellow circle marks its $\rtwo$. Credits: \citet{vazza18dynamo}/F. Vazza (top and central panels) and \citet{botteon22a2255} (bottom panel).}
 \label{fig:sim}
\end{figure}

Mergers between galaxy clusters are the most energetic events in Universe since the Big Bang, releasing up to 10$^{64}$ erg in the intracluster medium (ICM) on a time scale of a few Gyr. This energy is dissipated on very large scales by weak shocks and turbulence (Fig.~\ref{fig:sim}), mainly heating the thermal gas and partially accelerating relativistic particles and amplifying cluster magnetic fields \citep[\eg][for a review]{brunetti14rev}. These nonthermal components eventually manifest themselves as diffuse synchrotron emission in the ICM. Extended cluster-wide sources, called radio halos and relics, have been successfully detected in a large fraction of massive and merging clusters \citep[\eg][for a review]{vanweeren19rev}. With the advent of highly sensitive observations from \ska\ pathfinder and precursor instruments, we can now efficiently investigate how turbulent motions and shocks, predicted by simulations to pervade the ICM (Fig.~\ref{fig:sim}), dissipate kinetic energy into nonthermal components even at large distances from the cluster center.

\begin{figure*}[!t]
 \centering
 \includegraphics[width=\hsize,trim={0cm 0cm 0.3cm 0.28cm},clip]{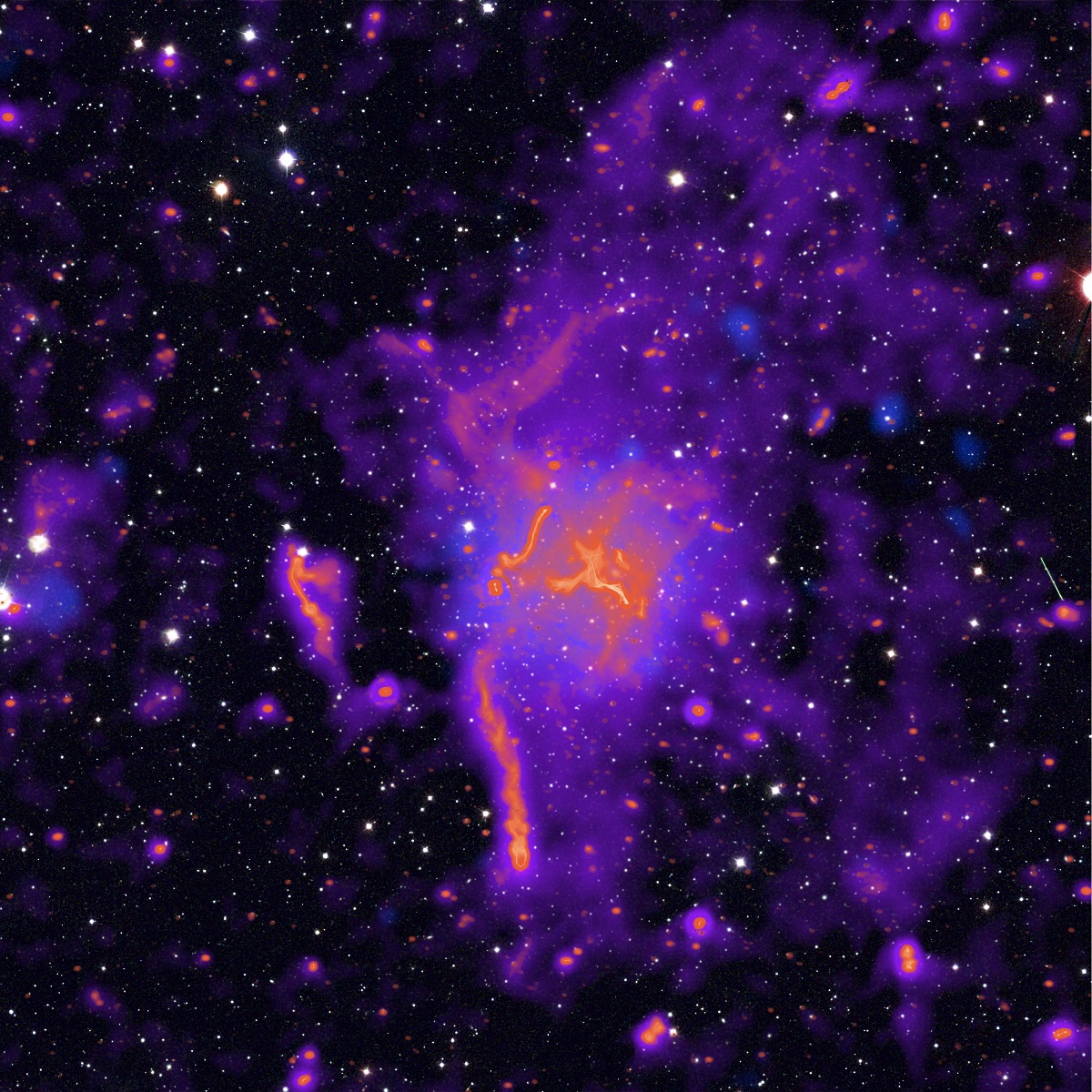}
 \caption{Composite image of Abell 2255, showing an area of about 1 deg$^2$ ($\sim$25 Mpc$^2$). X-ray data (blue) from \rosat\ shows the hot ICM; radio data (orange and purple) from LOFAR shows the nonthermal emission; the purple glow is the radio emission surrounding the entire cluster. Background optical image taken with the SDSS. Credits: \rosat/\lofar/\sdss/\citet{botteon22a2255}/F. Sweijen.}
 \label{fig:composite}
\end{figure*}

\subsection*{The LOFAR Galaxy Cluster Deep Field}

Abell 2255 is a massive cluster at $z=0.08$ undergoing a significant merger event, possibly involving multiple substructures colliding along different merger axes \citep[\eg][]{golovich19atlas}. Past radio observations show a rich variety of structures, from a central radio halo and a bright radio relic in the NE \citep{feretti97a2255, govoni05}, to various tailed radio galaxies \citep{harris80eight} and extended radio structures located in the cluster outskirts \citep{pizzo08, pizzo11, pizzo09}. Using 8~hr observations from the \lotssE\ \citep[\lotss;][]{shimwell17, shimwell19}, \citet{botteon20a2255} highlighted the multitude of interactions between head tail radio galaxies, the halo, and the ICM in the cluster center, discovering steep spectrum radio sources and narrow threads of emission embedded in the radio halo. \\
\indent
Due to its intricate environment, Abell 2255 is an ideal laboratory to study a plethora of different nonthermal phenomena in clusters across various scales. For this reason, it has been designated as the ``\lofar\ Galaxy Cluster Deep Field''. Between 2019 and 2021, Abell 2255 was observed for 72~h with \lofar-HBA (120--168 MHz) and for 72~h with \lofar-LBA (22--70 MHz). Below we discuss some of the key findings obtained from the analysis of these data, while we refer the reader to \citet{botteon22a2255} for a comprehensive description of the analysis, methods, and results. Between 2022 and 2023, additional 264~h of \lofar-HBA observations have been collected on the cluster. The analysis of these new observations is still ongoing.

\subsection*{Results}

The images produced from the 72~h LOFAR-HBA + 72~h LOFAR-LBA observations are the deepest ever obtained on a galaxy cluster at low frequencies. They reveal, in remarkable detail, pervasive radio emission that extends on much larger scales than the thermal ICM emission observed in X-rays, see Fig.~\ref{fig:composite}. This diffuse radio emission spans a largest projected linear scale of $\sim$5 Mpc, reaching out to the cluster virial radius, and is detected both with LOFAR-HBA and LOFAR-LBA. The extended ``envelope'' of low surface brightness radio emission (reported in purple in Fig.~\ref{fig:composite}) that engulfs the cluster central region implies the presence of relativistic electrons with energies of the order 1--10 GeV and magnetic fields at (sub-)\muG\ level distributed on very large scales, suggesting an efficient dissipation of kinetic energy into nonthermal components in the cluster outskirts. \\
\indent
Under the simple assumption of energy equipartition between magnetic fields and nonthermal electrons, we used the observed properties of the synchrotron radiation (namely, spectral index and luminosity) to estimate an equiparition magnetic field of

\begin{equation}\label{eq:eqi}
 B_{\rm eq} \simeq 0.45 (1+k)^{0.222} (\gamma_{\rm min}/1000)^{-0.409}\:\rm{\mu G}
\end{equation}

\noindent
at a distance of $\sim$2 Mpc from the cluster center (which roughly corresponds to \rtwo). In this equation, $k$ is the proton-to-electron energy ratio and $\gamma_{\rm min}$ is the minimum energy of relativistic electrons. Eq.~\ref{eq:eqi} implies an equipartition energy density of the nonthermal components of $\varepsilon_{\rm nth, eq} \sim 2 \times 10^{-14}$ erg cm$^{-3}$ and to ratio of nonthermal-to-thermal energy density of $\varepsilon_{\rm nth, eq}/\varepsilon_{\rm th} \sim 0.05$ ($\varepsilon_{\rm th}$ is derived from the pressure profile of Abell 2255 reported in \citealt{ghirardini19universal}). \\
\indent
To further refine our analysis, we can assume that the magnetic field $B$ deviates by a factor $\Delta$ from the value estimated in Eq.~\ref{eq:eqi}, \ie\ $B = B_{\rm eq} \Delta$. In this case, the magnetic field strength needs to fall within the range

\begin{equation}\label{eq:b_real}
 0.1 (\gamma_{\rm min}/1000)^{-0.409} \lesssim \left( \frac{B}{\muG} \right) \lesssim 1.7(\gamma_{\rm min}/1000)^{0.31}
\end{equation}

\noindent
in order to prevent that the energy content of the nonthermal components exceeds that of the thermal gas. Given that the radio emission we observe at low frequencies has steep spectrum, $\gamma_{\rm min}$ cannot be significantly higher than 1000. Lower values of $\gamma_{\rm min}$ would in turn reduce the range of possible $B$ values. For full derivations of Eqs.~\ref{eq:eqi} and \ref{eq:b_real}, we refer to \citet{botteon22a2255}.

\begin{figure}[t]
 \centering
 \includegraphics[width=\hsize,trim={0cm 0cm 0cm 0cm},clip]{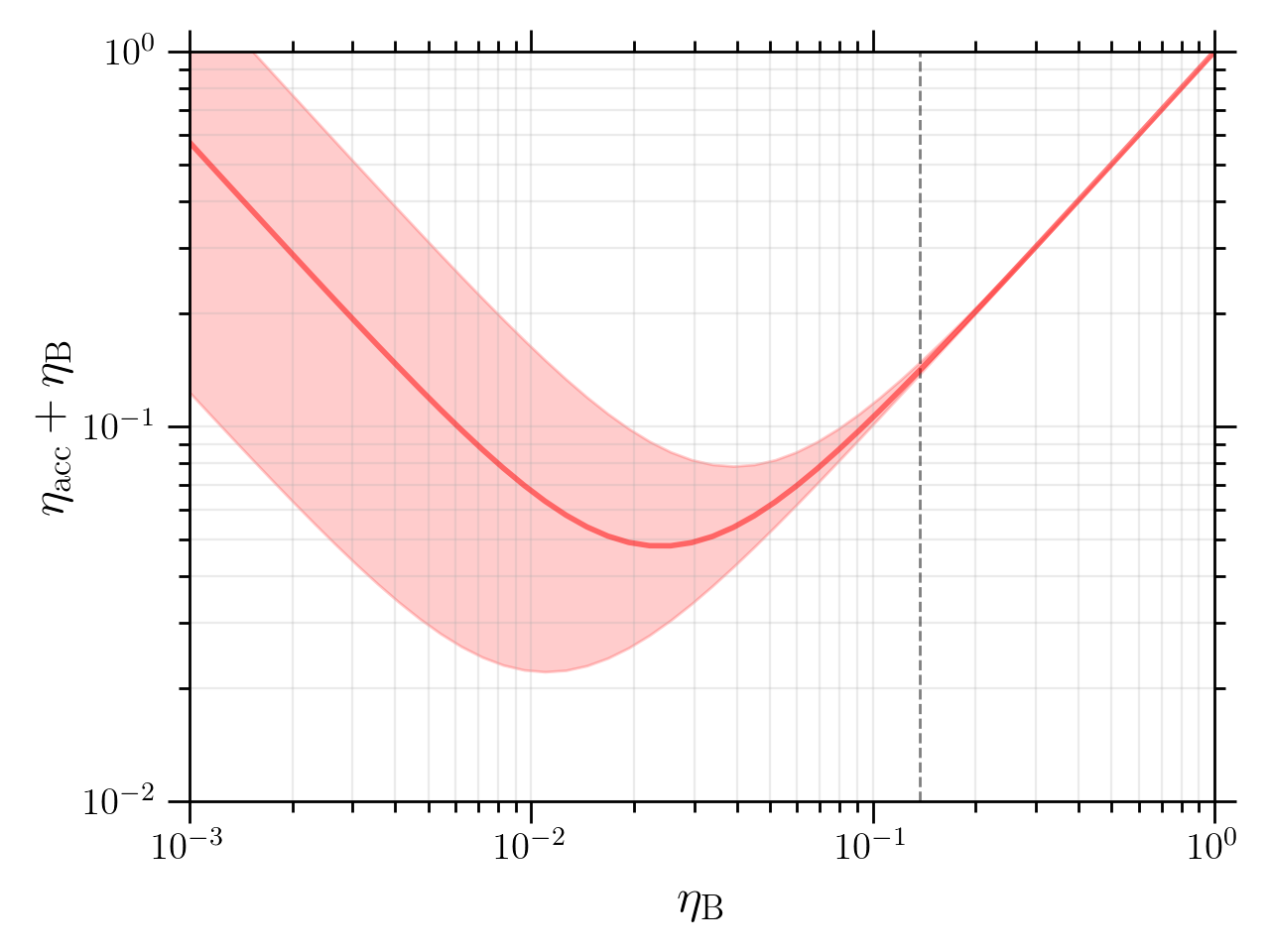}
 \caption{The solid line shows the fraction of turbulent energy that is converted into particle acceleration and magnetic field amplification ($\eta_{\rm acc}+\eta_{\rm B}$) as a function of that converted into amplification of the field only ($\eta_{\rm B}$). The dashed vertical line indicates the equipartition magnetic field $B_{\rm eq} = 0.45$ \muG. Adapted from \citet{botteon22a2255}.}
 \label{fig:eff}
\end{figure}

\begin{figure*}[t]
 \centering
 \includegraphics[width=\hsize,trim={1cm 7.5cm 1.5cm 5.2cm},clip]{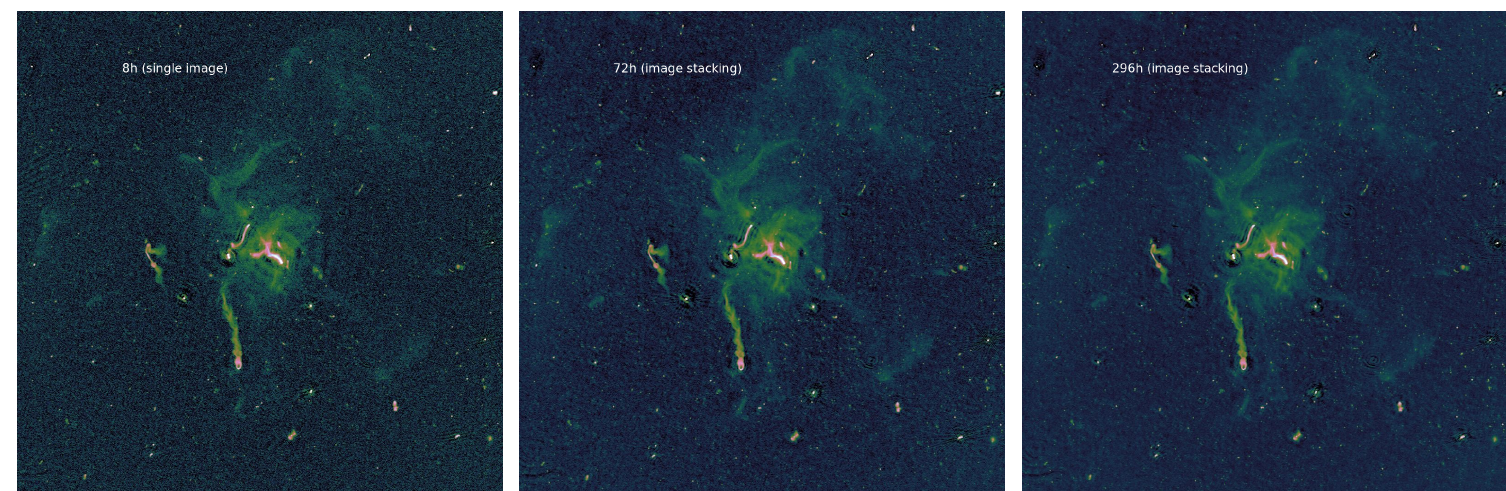}
 \caption{Comparison between \lofar-HBA images obtained with different amounts of data: a single 8~h observation (\textit{left}) and the stacked images from nine (\textit{center}) and thirty-seven (\textit{right}) 8~h observations.}
 \label{fig:stacking}
\end{figure*}

While the range 0.1--1.7 \muG\ is broad, it still provides valuable insights into the magnetic fields in cluster outskirts. The lower bound of 0.1~\muG\ that we estimated at $\sim$2 Mpc from the center of Abell 2255 is $\sim$250$\times$ the upper limit obtained on the primordial magnetic field using the cosmic background radiation \citep[\eg][]{paoletti19}. Moreover, it is at least 10$\times$ higher than field strength expected by the simple compression of a primordial field in the regions where we detect synchrotron emission. This suggests that the magnetic fields in the outskirts of galaxy clusters must be efficiently amplified from weak seed fields to reach $\sim$\muG-level strengths. Numerical simulations support this hypothesis, showing that small-scale turbulent dynamo amplification can produce such field strengths, as long as the effective Reynolds number in the ICM is sufficiently large \citep[\eg][]{beresnyak16turbulent, vazza18dynamo, dominguezfernandez19, xu20nonlinear}. \\
\indent
The detection of diffuse synchrotron emission at large cluster-centric radii enables the investigation of how the turbulent energy flux, $F_{\rm turb}$, is dissipated into nonthermal components in cluster outskirts. This occurs through two primary mechanisms: magnetic field amplification

\begin{equation}\label{eq:amp}
 \frac{B^2}{8\pi} \sim \eta_{\rm B} F_{\rm turb} \tau_{\rm eddy}
\end{equation}

\noindent
and particle acceleration

\begin{equation}\label{eq:acc}
 L_{\rm nth} \sim \eta_{\rm acc} F_{\rm turb} V
\end{equation}

\noindent
where $\eta_{\rm B}$ and $\eta_{\rm acc}$ are the efficiencies of the processes, $\tau_{\rm eddy}$ is the eddy turnover time of turbulence, and $V$ is the emitting volume where the accelerated electrons generate the nonthermal luminosity $L_{\rm nth}$ via synchrotron and inverse Compton emission. The total nonthermal luminosity is

\begin{equation}\label{eq:lnt}
 L_{\rm nth} = L_{\rm syn,bol} \left( 1 + \frac{B^2_{\rm ic}}{B^2} \right)
\end{equation}

\noindent
where $L_{\rm syn,bol}$ is the synchrotron bolometric luminosity and $B_{\rm ic}$ the equivalent inverse Compton magnetic field. By combining Eqs.~\ref{eq:amp}, \ref{eq:acc} and \ref{eq:lnt}, we can write the acceleration efficiency as a function of the magnetic field amplification efficiency as follows

\begin{equation}\label{eq:eff}
 \eta_{\rm acc} (\eta_{\rm B}) = \frac{L_{\rm syn,bol}}{F_{\rm turb} V} \left( 1 + \frac{B^2_{\rm ic}} {8\pi F_{\rm turb} \tau_{\rm eddy} \eta_{\rm B}} \right)
\end{equation}

\noindent
and use the results of numerical simulations of \citet{vazza18dynamo} to estimate $F_{\rm turb}$ and $\tau_{\rm eddy}$ in a simulated cluster resembling Abell 2255 (\cf\ Fig.~\ref{fig:sim}). Again, we refer the interested reader to \citet{botteon22a2255} for more details. By combining the values derived from observations and simulations, we produced the plot reported in Fig.~\ref{fig:eff} which indicates that 5 to 10\% of the turbulent energy flux needs to channeled into nonthermal components (particle acceleration and magnetic field amplification) in order to reproduce the extended radio emission detected with \lofar\ at 1.5--2 Mpc from the center of Abell 2255.

\subsection*{Conclusions and Perspectives}

The large-scale radio emission observed in Abell 2255 indicates that relativistic electrons and magnetic fields fill a significant fraction of the cluster volume and that are spread up to the virial radius. This allowed us to probe the properties of nonthermal components in poorly explored regions of the ICM and to provide estimates of the magnetic field strength and energy budget of nonthermal components at 1.5--2 Mpc from the cluster center. These results, derived from 72 h LOFAR-HBA + 72 h LOFAR-LBA observations, have been presented in detail in \citet{botteon22a2255}. \\
\indent
While the analysis of the full 336~h LOFAR-HBA observations of Abell 2255 is still ongoing, Fig.~\ref{fig:stacking} provides a preliminary look at the stacked image from thirty-seven 8~h observations, totaling 296~h of data. This image is presented alongside a single 8~h observation (equivalent to the amount of data presented in \citealt{botteon20a2255}) and the stacked image from nine 8~h observations (equivalent to the amount of data presented in \citealt{botteon22a2255}). We are currently in the phase of optimizing the direction-dependent calibration of the ultra-deep dataset. Once completed, the diffuse emission engulfing the cluster will be detected at higher significance, and additional faint emission may be recovered on broader scales. Whilst polarization studies of ICM magnetic fields at the low frequencies where \lofar\ operates are particularly challenging due to significant Faraday depolarization effects, the ultra-deep data and detection of synchrotron emission in low-density environments, where depolarization is less severe, may provide the possibility to perform this kind of analysis in Abell 2255. Alternatively, a method such as the synchrotron intensity gradient \citep{hu24} technique, which operates on total intensity images, may be employed to infer the magnetic field orientation underlying the detected diffuse radio emission.

\vspace*{5mm}
\noindent
{\scriptsize {\bf Acknowledgments.} A.B. acknowledges financial support from the European Union - Next Generation EU.}

\newpage

\bibliographystyle{aa-only-first-author}
\bibliography{library.bib}

\end{document}